\documentclass[12pt]{iopart}
\usepackage{graphicx,iopams}
\usepackage{amsfonts}
\newcommand{\EQ}{\begin{equation}}
\newcommand{\EN}{\end{equation}}
\newcommand{\be}{\begin{equation}}
\newcommand{\ee}{\end{equation}}
\newcommand{\bea}{\begin{eqnarray}}
\newcommand{\eea}{\end{eqnarray}}
\def\th{\theta}

\begin{document}
\bibliographystyle{unsrt}  
\title[Two-kink bound states in the Potts field theory]
{Two-kink bound states in the magnetically perturbed Potts field theory at $T<T_c$}
\author{S.~B. Rutkevich}
\address{Institute of Solid State and Semiconductor Physics,  SSPA
"Scientific-Practical Materials Research Centre,
NAS of Belarus", P.~Brovka St. 17, 220072 Minsk, Belarus}
\begin{abstract}
The $q$-state Potts field theory with $2\le q\le 4$ in the low-temperature phase is considered in presence of a 
weak magnetic field $h$. In absence of the magnetic field, the theory is integrable, but not free at $q>2$: its elementary
 excitations - the kinks - interact at small distances, and their interaction can be characterized by the
  factorizable scattering matrix  which was found by Chim and Zamolodchikov.
The magnetic field induces the long-range attraction between kinks causing  their
confinement into the bound-states.
We calculate the masses of the two-kink bound states in the leading order in $h\to \pm0$ expressing them in terms
of the  scattering matrix of kinks at $h=0$.		
\end{abstract}
\ead{rut@ifttp.bas-net.by, rut57@mail.ru}
\section{Introduction}
The kink topological excitations are quite common in the two dimensional field theories with
Hamiltonian invariant under some discrete symmetry group $\mathcal G$. If such a symmetry is spontaneously broken
in the ordered phase, the latter has a discrete set of degenerate vacua $|0_\alpha\rangle$, $\alpha=1,\ldots,q$.
Then the kinks $K_{\alpha\beta}$, i.e. the  domain walls separating two different vacua $\alpha$ and $\beta$,
behave like stable quantum particles which can propagate in the system.
Adding a small interaction, which explicitly breaks the Hamiltonian $\mathcal G$-symmetry, lifts the degeneracy
of ground states $|0_\alpha\rangle$ and leads to confinement of kinks dividing the true and false vacua.
This simple, but quite general scenario of confinement in two dimension originates to the work of
McCoy and Wu \cite{McCoy78}.
Its particular realizations  in different two-dimensional
models have been the subject of a considerable interest in the recent years \cite{FZ06,Tsv04,DelMus98,Del08,Mus07,
Mus08,MusTak09}.

The simplest and most studied example of the two-dimensional model exhibiting confinement is the
Ising Field Theory  (IFT) \cite{FZ06}, which is characterized by the $\mathbb{Z}_2$ symmetry group.
 At zero magnetic field, the theory describes  free massive (apart from the critical point)
neutral fermions, which represent in the low temperature phase the kinks interpolating between two degenerate vacua.
At small magnetic field $h>0$, the  kinks ('quarks') become confined into pairs which form a tower of bound states
('mesons') having zero topological charge. The meson masses $M_n(h)$
 densely fill the segment{\footnote {Note, that only the mesons having masses $ M_n(h)$ below the 
 two masses of the lightest meson are the stable particles: $M_n(h)<M_1(h)\approx 4m$. Heavier mesons are the resonances
since they can decay into a pair of light mesons.}}
 $[2m,\infty)$ at $h\to 0$. Two  asymptotic expansions describe  $M_n(h)$
at small $h$ in different regions of this segment.
\begin{enumerate}
\item
Near the  edge point $2m$ (i.e. for
fixed $n$ at $h\to0$) one can use the {\it low energy expansion} in fractional powers
of the magnetic field. Its leading term was obtained by McCoy and Wu \cite{McCoy78},
further corrections were found by Fonseca and Alexander Zamolodchikov \cite{FZ06,FonZam2003}.
\item
The masses of highly excited mesons
(with $n \gg 1$, in particular for $n \sim 1/h$) are described by the
{\it semiclassical expansion} in the integer powers of $h$  \cite{ FZ06,Rut05,Rut09}.
\end{enumerate}

The both low-energy and semiclassical expansions for the meson masses in the IFT
 were obtained by means of
the perturbative analysis of the Bethe-Salpeter equation, first derived for
this model by Fonseca and  Zamolodchikov \cite{FonZam2003}.
Since their original derivation  procedure substantially exploited the
free-fermionic structure of the Ising field theory at $h=0$, it was not applied to
the models, in which kinks interact
at short distances already in the deconfined phase.

In this paper we address to the problem  of extension of the Bethe-Salpeter approach
to make it appropriate
for calculation of meson masses in such two-dimensional models.
The particular subject of our interest is  the $q$-state Potts Field Theory (PFT) for $2\le q\le4$,
which describes the scaling limit of the two-dimensional Potts model. In the latter,
each site in the lattice has $q$ different states ('colours'). At zero magnetic field, the Potts model
is invariant under the group $S_q$ of permutations of $q$ colours. At $q=2$, the Potts model
reduces to the Ising model.

The zero-field PFT is integrable, i.e. it has infinite number of integrals
of motion and the factorizable scattering matrix \cite{CZ92}. In the low temperature phase,
the particle content of the (zero-field ) PFT contains $q(q-1)$ kinks $K_{\alpha\beta}$,
$\alpha,\beta=1,\ldots,q$, which interact with each other at short distances. Their $S$-matrix
was found by Chim and Zamolodchikov \cite{CZ92}.

The PFT in presence of a nonzero
magnetic field $h$ acting on one of the $q$ colours has been studied recently by Delfino and Grinza \cite{Del08}.
Beyond other result relating to the $T>T_c$ phase, these authors performed qualitative analysis
of the kink confinement in this model, classified two-kink and three kink-bound states,
and conjectured evolution of their mass spectra with temperature and magnetic field.

The main subject of our interest are the meson masses $M_n(h)$  in the $q$-state PFT
in the low temperature phase in the limit of the weak magnetic field $h\to 0$. As in reference \cite{Del08},
the magnetic field is chosen acting on one colour only.
We consider the mesons as the
bound states of two kinks, which attract one another with a linear potential at large distances,
and undergo scattering upon collisions. As the result, we express the leading terms in the both low-energy
and semiclassical expansions for $M_n(h)$ in terms of the known zero-field kink $S$-matrix.

The paper is organized as follows.  In the next section, we remind the definition and some well known
properties
of the Potts model on the square lattice and its scaling limit. Sections \ref{Sect3} and \ref{hneg}
describe calculation of the meson masses in the leading order of the weak magnetic field
in the low temperature phase at $h\to+0$ and at $h\to -0$, respectively, for different values of $q$.
In Section \ref{sect:BS} the Bethe-Salpeter equation for the PFT is derived. In the IFT, the 
analogous Bethe-Salpeter equation provides the basis for the systematic theory of mesons \cite{FZ06}.  
Concluding remarks are presented in section \ref{Conc}.

\section{The Potts model and its scaling limit \label{Pm}}
In this section we describe briefly the definition and few basic properties
of the $q$-state Potts model in two dimensions. We start from the square
lattice model and then pass to its scaling limit.

Consider the two-dimensional  square lattice ${\mathbb Z}^2$,
and associate the discrete spin variable
$s(x) =1,2,\ldots,q$  with each lattice site $x\in {\mathbb Z}^2 $. The model
Hamiltonian is  defined as
\begin{equation}
{\cal H}=-\frac{1}{T} \sum_{<x,\,y>} \delta_{s(x),s(y)}-H\sum_x
  \delta_{s(x),q}.
\label{Ham}
\end{equation}
Here the first summation is over the nearest neighbour pairs, $T$ is the
temperature, $H$ is the external magnetic field
applied along the $q$-th direction, and $\delta_{\alpha,\alpha'}$ is the Kronecker symbol.
 At $H=0$, Hamiltonian (\ref{Ham}) is invariant under the permutation group $S_{q}$, at $H\ne0$
the symmetry group reduces to $S_{q-1}$. By means of mapping  onto the random cluster model \cite{KasFor69,KasFor72}, one can
also define the $q$-state Potts model with noninteger values of $q$.

The order parameters $\langle\sigma_\alpha\rangle$ can be associated with the variables
\[
\sigma_\alpha(x) =\delta_{s(x),\alpha}-\frac{1}{q}, \quad \alpha=1,\ldots,q.
\]
Parameters $\langle\sigma_\alpha\rangle$ are not independent, since
\begin{equation}
\sum_{\alpha=1}^q\sigma_\alpha(x)=0.
\end{equation}
The zero-field model undergoes ferromagnetic phase transition at the critical temperature
\begin{equation}
  T_c=\frac{1}{ \log(1+\sqrt{q})}.
\end{equation}
This phase transition is first order for $q>4$, and continuous for $2\le q\le 4$.
The ferromagnetic low-temperature phase at zero field
is $q$-times degenerated. 
For a review of many others known properties of the Potts model see \cite{Bax,Wu82}.

We shall consider only the Potts model with $2\le q\le 4$.
In this case, the correlation length diverges at $H\to 0$, $T\to T_c$.
The Conformal Field Theory  associated with this
critical point is characterized by the central charge
\begin{eqnarray}
c(q)=1-\frac{6}{t(t+1)}, \quad {\rm where}
\quad \sqrt{q}=2 \sin\frac{\pi(t-1)}{2(t+1)}.
\end{eqnarray}
The scaling limit of
the model (\ref{Ham}) is described by the action \cite{Del08}
\begin{equation} \label{AP}
{\cal A}= {\cal A}_{CFT}^{(q)} -\tau\int \rmd^2x\,\varepsilon(x)-
h\int \rmd^2x\,\sigma_q(x)\,\,,
\label{scaling}
\end{equation}
Here ${\cal A}_{CFT}^{(q)}$ corresponds to the Conformal Field Theory,
which is associated with the the critical point. Fields $\varepsilon(x)$ (energy density)
and $\sigma_q(x)$ (spin density) have the
scaling dimensions
\[
X_\epsilon^{(q)}=\frac{1}{2}\left(1+\frac{3}{t}\right),  \quad \quad
X_\sigma^{(q)}=\frac{(t-1)(t+3)}{8 t(t+1)}.
\]
Couplings $\tau$ and $h$ are proportional
to the deviation of the temperature and magnetic field
from their critical point values.
\subsection{Ordered phase in the PFT at $h=0$}
The field theory (\ref{AP}) is integrable along the line $h=0$ in the $(\tau,h)$-plane.
In this paper only the low-temperature ($\tau<0$) phase will be considered.
At $h=0$ and $\tau<0$,
the $S_q$ symmetry is  spontaneously broken: the model has $q$ degenerate
vacua $|0_\alpha\rangle$, $\alpha=1,2,\ldots,q$, which are distinguished
by the values of the order parameter
\begin{equation}
\langle \sigma_\gamma\rangle_\alpha \equiv \langle 0_\alpha|\sigma_\gamma(x)|0_\alpha\rangle=
\frac{v}{q-1}\left(q\,\delta_{\gamma,\alpha}-1\right),
\end{equation}
with some positive $v$.
The symmetry group $S_q$ acts by permutations of these vacua.
Elementary excitations are the $q(q-1)$ kinks $K_{\alpha \beta}(\theta)$, which interpolate between
different vacua $\alpha$ and $\beta$. Here $\theta$ denotes the kink rapidity, which  parametrizes
its energy and momentum
\begin{equation}
E=m\cosh \theta,\quad p=m \sinh \theta,  \label{Mn1}
\end{equation}
with $m \sim |\tau|^{1/[2-X_\epsilon^{(q)}]}$   being the kink mass.

The two-kink scattering at $\tau<0$, $h=0$ is described by the Faddeev-Zamolodchikov
commutation relations
\bea  \label{FZ}
K_{\alpha\gamma}(\theta_1)K_{\gamma\beta}(\theta_2) =\sum_{\delta\ne\alpha,\beta} S_{\alpha\beta}^{\gamma \delta}(\theta_{1,2})
K_{\alpha\delta}(\theta_2)K_{\delta\beta}(\theta_1), 
\eea
with the scattering amplitudes $S_{\alpha\beta}^{\gamma \delta}(\theta_{1,2})$, and $\theta_{12}= \theta_{1}-\theta_{2}$.
Due to the $S_q$ invariance, only four scattering amplitudes are independent, providing  
\bea
\fl K_{\alpha\gamma}(\theta_1)K_{\gamma\beta}(\theta_2) =S_0(\theta_{12})
\sum_{\delta\neq\gamma}K_{\alpha\delta}(\theta_2)K_{\delta\beta}(\theta_1)+
S_1(\theta_{12})
K_{\alpha\gamma}(\theta_2)K_{\gamma\beta}(\th_1)\,,\hspace{.5cm}\alpha\neq\beta
\label{KSc}\\
\fl K_{\alpha\gamma}(\theta_1)K_{\gamma\alpha}(\theta_2)=S_2(\theta_{12})
\sum_{\delta\neq\gamma}K_{\alpha\delta}(\theta_2)K_{\delta\alpha}(\theta_1)+
S_3(\theta_{12})K_{\alpha\gamma}(\theta_2)K_{\gamma\alpha}(\theta_1)\,\,, \label{KS}
\eea
The explicit expressions for the  scattering amplitudes
were determined in reference  \cite{CZ92}:
\bea
&& S_0(\theta)=\frac{\sinh\lambda\theta\,\sinh\lambda(\theta-\rmi\pi)}
{\sinh\lambda\left(\theta-\frac{2\pi \rmi }{3}\right)\,\sinh\lambda\left(\theta-
\frac{\rmi \pi}{3}\right)}\,\Pi\left(\frac{\lambda\theta}{\rmi \pi}\right)\,,
\label{s0}\\
&& S_1(\theta)=\frac{\sin\frac{2\pi\lambda}{3}\,\sinh\lambda(\theta-\rmi \pi)}
{\sin\frac{\pi\lambda}{3}\,\sinh\lambda\left(\theta-\frac{2 \rmi \pi}{3}\right)}\,
\Pi\left(\frac{\lambda\theta}{\rmi \pi}\right)\,, \label{S1}\\
&& S_2(\theta)=\frac{\sin\frac{2\pi\lambda}{3}\,\sinh\lambda\theta}
{\sin\frac{\pi\lambda}{3}\,\sinh\lambda\left(\theta-\frac{\rmi \pi}{3}\right)}\,
\Pi\left(\frac{\lambda\theta}{\rmi \pi}\right)\,,\label{s2}\\
&& S_3(\theta)=\frac{\sin\lambda\pi}{\sin\frac{\pi\lambda}{3}}\,
\Pi\left(\frac{\lambda\theta}{\rmi \pi}\right)\,.\label{s3}
\eea
Parameter $\lambda$ is related to $q$ as
\EQ
\sqrt{q}=2\sin\frac{\pi\lambda}{3}\,,
\label{qlambda}
\EN
and
\bea
&&\Pi\left(\frac{\lambda\theta}{\rmi\pi}\right)=
\frac{\sinh\lambda\left(\theta+\rmi\frac{\pi}{3}\right)}{\sinh\lambda(\theta-
\rmi\pi)}\,e^{{\cal A}(\theta)}\,,\\
&& {\cal A}(\theta)=\int_0^\infty\frac{\rmd x}{x}\,
\frac{\sinh\frac{x}{2}\left(1-\frac{1}{\lambda}\right)-
\sinh\frac{x}{2}\left(\frac{1}{\lambda}-\frac{5}{3}\right)}
{\sinh\frac{x}{2\lambda}\cosh\frac{x}{2}}\,\sinh\frac{x\theta}{\rmi\pi}\,\,. \label{Aex}
\eea
Note, that the values of the parameter $\lambda$ corresponding to integer $q$ are:
\begin{eqnarray*}
\lambda=\frac{3}{4} \quad {\rm for} \; q=2,\\\
\lambda=1 \quad {\rm for} \; q=3,\\
\lambda=\frac{3}{2} \quad {\rm for} \; q=4.
\end{eqnarray*}
\subsection{Kink confinement in a weak magnetic field}
Application of a small magnetic field along the $q$-direction lifts degeneracy between the
vacuum $|0_q\rangle$ and vacua $|0_\alpha\rangle$ with $\alpha<q$.
In the first order in $h$, the  shift $\Delta {\mathcal E} $ between their energy densities reads as
\begin{equation} \label{str}
\Delta {\mathcal E} =\delta{\mathcal E}_\alpha-\delta{\mathcal E}_q= \frac{v q}{q-1}\, h,
\quad {\rm for }\quad\alpha=1,\ldots,q-1\,\,.
\end{equation}
It gives rise to the linear attractive potential between two kinks which interpolate between the stable and false
vacua,
\be  \label{attr}
V(x_1,x_2)=(x_2-x_1) \,\Delta {\cal E}
\ee
where $x_1<x_2$ are the spacial coordinates of the  kinks.

Depending on the sign of $h$, two regimes are distinguished \cite{Del08}.
\begin{itemize}
\item
If $h>0$, the vacuum $|0_q \rangle$ becomes the true ground state of the system, and
the states $|0_\alpha \rangle$ with $\alpha\ne q$ become the false vacua. Magnetic field induces
a long-range attraction between kinks leading to their confinement. Isolated kinks do not survive as asymptotic states
of the theory, and the elementary excitations are the bound states of two and three  kinks.
\footnote {The three-kink bound states can exist if $3\le q\le 4$.  The four-kink bound states, which are allowed
 for $q=4$, should be unstable due to their decay into a pair of mesons \cite{Del08}. }
\item
If $h<0$, the vacuum $|0_q \rangle$ becomes metastable, and the true vacuum
states $|0_\alpha \rangle$ with $\alpha=1,\ldots,q-1$
are still degenerate in the energy. Elementary excitations are the kinks $K_{\alpha \beta}(\theta)$ interpolating
between the true vacua $\alpha,\beta\ne q$. On the other hand, the kinks
$K_{\alpha q}(\theta_1)$ and $K_{q\beta}(\theta_2)$
are confined into the bound states by the magnetic field.
However, such bound states are  unstable due to decay into isolated kinks $K_{\alpha \beta}(\theta)$
(see \cite{Del08} and the discussion below in Section \ref{hneg}).
\end{itemize}
Note, the kinks in the fields theories with confinement are often called 'quarks', while their
bound states play the role of 'mesons' (kink-antikink states), and 'baryons' (tree-kink states).
\section{Meson masses at $h\to+0$ \label{Sect3}}
The meson mass $M$ can be formally determined from the solution of the
eigenvalue problem
\bea \label{en}
\hat{\mathcal H}(h) \,|\pi(P)\rangle =\left[E(P) +E_{vac} \right]\,|\pi(P)\rangle ,\\
\hat{P} \,|\pi(P)\rangle = P\,|\pi(P)\rangle,  \label{mom}
\eea
where $\hat{\mathcal H}(h)$ is the Hamiltonian,  and $\hat{P}$ is the total momentum operator
corresponding to action (\ref{AP}), $E_{vac}$ is the vacuum energy, $E(P)$
is the meson energy spectrum, which should have the relativistic form
\be
E(P)=(P^2+M^2)^{1/2}.  \label{relM}
\ee
 Unfortunately, the explicit form of the PFT Hamiltonian
$\hat{\mathcal H}(h)$ is known only in the case $q=2$, which corresponds  to the Ising Field Theory. Below we
describe briefly, how the meson masses at $q=2$ can be calculated from the perturbative analysis of the Bethe-Salpeter
equation in the coordinate representation \cite{FonZam2003,Rut05}. This procedure will be then naturally
generalized  to the  case $2< q\le 4$.
\subsection{Ising Field Theory case: $q=2$}
In the IFT, kinks are  fermions which are free at $h=0$. For nonzero $h$, the  Hamiltonian $\hat{\mathcal H}(h)$
can be written as
\begin{eqnarray}
\hat{\mathcal{H}}(h)= \int_{-\infty}^\infty \frac{\rmd p}{2 \pi} \,\omega(p)\, a^{\dagger}_p\, a_p
-h\int_{-\infty}^\infty  \rmd x\,\sigma_2(x),   \label{V}
\end{eqnarray}
where
$\omega(p)=\sqrt{p^2+m^2}$ is the spectrum of free fermions, and formfactors of the spin operator
$\sigma_2({x})$   are explicitly known \cite{Berg79}. At small $h$,  one can treat the meson as a
bound state of two quarks neglecting four-quark, six-quark, etc. contributions in its wave function.
This two-quark approximation is asymptotically exact in the leading order in $h\to0$.
In this approximation, the meson mass $M_n$ can be calculated from the
perturbative solution of  the Bethe-Salpeter equation.
In the coordinate representation, the latter equation written in the meson rest frame reads as
\begin{eqnarray}
2 \omega(\hat{p})\, \phi^{(n)} ({ x})+  \Delta {\mathcal E}
 |{x}|\,\phi^{(n)}({x}) + \Delta {\mathcal E}  \,\hat{U} \, \phi^{(n)} ({x})
= M_n\, \phi^{(n)} ( x) , \label{BSx} \\
{\rm where}\quad -\infty<x<\infty.  \nonumber
\end{eqnarray}
Here $\Delta {\mathcal E}=2 h\langle \sigma_2\rangle$ is the 'string tension',
$\langle\sigma_2\rangle $ is the
spontaneous magnetization at zero field, $|x|$ is the distance between the two quarks,
$\hat{p}=-i\partial_x$,  and $ \phi^{(n)} ({ x})$ denotes the configuration-space wave function
in the two-quark approximation
\begin{eqnarray}
\phi^{(n)} ({x})= \int_{-\infty} ^\infty\frac{\rmd p}{2 \pi} \,e^{\rmi p { x}}
\langle0|a_{(P-p)}a_p|\pi_n(P)\rangle_{P\to0}\,\,,   \label{Fou}\\
\phi^{(n)} (-{x})=- \phi^{(n)} ({ x}). \label{odd}
\end{eqnarray}
The first and the second terms in the left-hand side of (\ref{BSx}) correspond to the kinetic energy of two quarks,
and to their long-range attraction, respectively.  The linear integral operator $  \hat{U} $ describes the
 the short-range
interaction between quarks in the two-fermion sector,
its kernel $  {U}(x,x') $   exponentially vanishes at large distances
$|{x}|\gg m^{-1} $.

Several perturbative schemes have been developed for the IFT Bethe-Salpeter equation.
The most convenient for us is the procedure described in reference \cite{Rut05}. It is based on the
observation, that at small $h$, the average distance between the quarks in the meson  is
large compared with the correlation length $m^{-1}$. Therefore, in the
right 'transport region' at
$x\gg m^{-1}$, one can very well approximate the solution $\phi_{n}(x)$
of the Bethe-Salpeter equation (\ref{BSx}) by the bounded
in the whole line $-\infty<x<\infty$ solution $ \Phi^{(n)} (x) $ of the equation
\be\label{qc}
\left[2 \,\omega(\hat{p})-M_n+   \Delta {\mathcal E}  \, x\right]\, \Phi^{(n)}({x})= 0.
\ee
After the Fourier transform this equation takes the form
\be
\left[ 2\,\omega(p)  -M_n+\rmi\, \Delta {\cal E} \,\partial_p \right] \, \Phi^{(n)}(p) =0,
\ee
providing
\be
\Phi^{(n)} (x)= \int_{\infty}^{\infty}\frac{\rmd p}{2\pi}\,
\exp \left\{\frac{\rmi[f(p)-p\, M_n]}{ \Delta {\cal E} }+\rmi p x\right\}  \label{Phi}
\ee
where
\be  \label{fdef}
f(p)=2\int_0^p \rmd q \,\omega(q)=m^2\left[\theta+\frac{\sinh 2\theta}{2}\right],
\ee
and $p=m\,\sinh \theta$.

In the left 'transport region' $x<0$, $|x|\gg m^{-1}$, function $\phi^{(n)}(x)$ should approach to
$- \Phi^{(n)} (-x)$ due to (\ref{odd}). In the intermediate "scattering region" $|x|\lesssim m^{-1}$,
one can solve equation (\ref{BSx}) perturbatively in $h$. Joining solutions obtained in these three
regions leads to the condition which gives the meson mass spectrum $M_n(h)$.

It turns out, that {\it in the leading order in } $h$, the meson masses can be  obtained from the
equation
\be
\Phi^{(n)} (x) |_{x=0}=0,  \label{Phi0}
\ee
since the odd continuous function $\phi^{(n)}(x)$ can be approximated (in the leading order in $h$)
by integral (\ref{Phi}) in the whole positive half-axis $x>0$.

At $x=0$, $M_n>2$ and $h\to+0$, the
integral in (\ref{Phi}) is determined by two saddle points $\pm p_n$, where
\begin{equation}
 p_n= m\sinh \beta_n\,,  \label{sadpo}
\end{equation}
and $\beta_n$ parametrizes the meson masses $M_n$:
\be
M_n=2 m \cosh \beta_n\,.   \label{Mn}
\ee
 In the leading order in $h$ this yields the first term of
the semiclassical expansion \cite{FZ06, Rut05}
\be   \label{Mth}
\sinh(2\beta_{n})-2\,\beta_{n}=
2\pi \left(n-\frac{1}{4}\right) \zeta +O(\zeta^2),
\ee
where $\zeta=\Delta {\cal E}/m^2\sim h$. The semiclassical expansion holds if $\zeta\ll 1$ and $n\gg1$.

If $M_n$ approaches  $2 \, m$, two saddle points $\pm p_n$ merge at the origin, and  integral (\ref{Phi})
becomes proportional to the Airy function
\bea
\Phi^{(n)} (x)\Big|_{x=0}=m\, \int_{-\infty}^\infty \frac{\rmd \theta}{2\pi} \exp\bigg\{\frac{\rmi}{ \zeta }\nonumber
\bigg[\frac{\theta^3}{3}-\theta\,\frac{(M_n-2 m)}{m}  \bigg]\bigg\}+O(\zeta)=\\
m\,\zeta^{1/3}
{\rm Ai}\left[-\,\frac{(M-2 m)}{m \,\zeta^{2/3} } \right]+O(\zeta) \label{Airy}
\eea
in the limit $h\to 0$. Formulae (\ref{Phi0}), (\ref{Airy}) give rise to the leading term
of the low-energy expansion \cite{McCoy78}
\be   \label{lMk}
\frac{M_n}{m}-2= z_n \,\zeta^{2/3}+O(\zeta^{4/3}),
\ee
where $-z_n$ denotes the $n$-th zero of the Airy function, ${\rm Ai}(-z_n)=0$, $n=1,2,\ldots$.
The low-energy expansion holds if  $n \zeta \ll 1$. It corresponds to the nonrelativistic approximation
in the quark dispersion law
$\omega ( \hat{p} )= m+\hat{p}^{\,2}/(2 m)+\ldots$
in equation (\ref{qc}).
\subsection{Generic $q$ \label{Genq}}
Let us discuss now, how the procedure described above should be modified to be applied
to  the PFT with $2<q\leq 4$, or more generally, to models exhibiting weak
confinement at small $h$, which are integrable but not free at $h=0$. In this Subsection
parameter $q$ is allowed to take fractional values, which are assumed to be not too close to 3.

First, it is natural to expect that the two-quark approximation can be safely used at small $h$,
if the distance between quarks is much larger than the correlation length $m^{-1}$. We adopt
two further assumptions.
\begin{enumerate}
\item \label{i}
At large distances $x_2-x_1\gg m^{-1}$, interaction between two quarks $K_{q\alpha}(x_1)$ and $K_{\alpha q}(x_2)$
forming a meson is completely described by the linear attractive potential
\be
V(x_1,x_2)= (x_2-x_1)\,\Delta{\mathcal E}   \label{Vi}
\ee
with the string tension
\be
\Delta{\mathcal E}= \frac{v q}{q-1} h+o(h).
\ee
\item  \label{ii}
To the linear order in the magnetic field, the quark dispersion law $\epsilon(p;h)$ is the same
as the free fermion spectrum $\omega(p)=(p^2+m^2)^{1/2}$\,,
\be
\epsilon(p;h)=\omega(p)+o(h). \label{eii}
\ee
\end{enumerate}

These natural assumptions summarize the experience gained from the IFT 
where they can be verified by means of the consistent perturbation theory 
based on the Bethe-Salpeter equation \cite{FZ06, Rut09}.  In particular, 
radiative corrections to the quark string tension and dispersion law in 
the IFT are known to be of the third, and second order in $h$, respectively. 
Furthermore, the approach based on the above assumptions 
allows one to reproduce in a simple way the leading order of the exact asymptotical 
expansions for the meson masses in the IFT, see Section 2 of reference \cite{FZ06}.

The Bethe-Salpeter equation (\ref{BSPotts}) for the $q$-state PFT will be derived in Section 
\ref{sect:BS}. The kernel of this integral equation is expressed in the matrix 
element (\ref{ME}) of the magnetization operator between the two-quark states with definite rapidities.
Since the explicit form of this matrix element  is not known for $q>2$, a
direct proof of statements (\ref{i}), (\ref{ii}) for general PFT is still impossible, 
and we shall take them as assumptions.

A zero-momentum meson state $\pi(0)$ can be characterized by the $(q-1)$-component
wave function $\psi_\alpha (x)$,
\be \label{psi}
\psi_\alpha (x) =\langle K_{q\alpha}(x_1+x)K_{\alpha q}(x_1) \mid \pi(0) \rangle,
\quad \alpha=1,\ldots,q-1,
\ee
which is well defined for large positive  $x\gg m^{-1}$.

Action (\ref{scaling}) and the true vacuum $|0_q\rangle$ are invariant under the group
$S_{q-1}$ of permutations of the first $q-1$ colours.
In what follows, we shall use the results of the $S_{q-1}$ symmetry analysis of the
meson states, which has been done by Delfino and Grinza \cite{Del08}.
The meson states form $(q-1)$ multiplets $ \pi_k $. Here  $ \pi_k $ are the
eigenstates of the generator
$ \Omega_{q-1} $ of cyclic permutations of first  $q-1$ colours $\alpha=1,\ldots,q-1$,
\bea
\Omega_{q-1} \pi_k= \gamma^k \,\pi_k,\\
{\rm where} \quad  \gamma=\exp[2\pi \rmi/(q-1)], \quad k=0,\ldots,q-2. \nonumber
\eea
This symmetry determines  the $\alpha$-dependence of the wave function $\psi_{k,\alpha} (x) $
for  the state $ \pi_k(0)$,
\be
\psi_{k,\,\alpha} (x)  = \gamma^{-k\,\alpha}  \phi_k (x).
\ee
Due to (\ref{Vi}) and (\ref{eii}), the meson wave function $\phi_k^{(n)}(x)$ corresponding to
the state $\pi_k^{(n)}(0)$ should satisfy
equation (\ref{qc}) at $x\gg m^{-1}$, and  therefore, we can write  it in the form
\be  \label{pphi}
\phi_k^{(n)}(x)\approx \int_{\infty}^{\infty}\frac{\rmd p}{2\pi}\,
\exp \left\{\frac{\rmi[f(p)-p\, M_{n}]}{ \Delta {\cal E} }+\rmi p x\right\} \quad{\rm for}\quad x\gg m^{-1},
\ee
where $f(p)$ is given by (\ref{fdef}).

 The further analysis depends on the value of $n$.
\subsubsection{Semiclassical case.}
For  large $n\gg 1$ and fixed $x\gtrsim m^{-1}$,    the $h\to 0$ asymptotics of integral (\ref{pphi})
is determined by contributions of two saddle-points (\ref{sadpo}), providing
\bea
\phi_k^{(n)}(x)\approx \left(\frac{ \Delta {\cal E} }{4 \pi \tanh \beta_n}\right)^{1/2}
\left[ C_k(\beta_n) e^{\rmi m x\sinh\beta_n }+ C_k^*(\beta_n) e^{-\rmi m x\sinh\beta_n }
\right] , \nonumber\\
 \label{Ck}
{\rm with }\quad C_k(\beta_n)=\exp \left[ \frac{\rmi \left(\beta_n-\frac{\sinh 2\beta_n}{2}\right)}
{\zeta}+ \frac{\rmi\pi}{4}\right].
\eea
Here again the parametrization (\ref{Mn})
is implied.
Coefficients  $C_k^*(\beta_n)$ and $C_k(\beta_n)$ are just the  in- and out- amplitudes,
which characterize the quark "plane waves" which  enter and leave
 at $x\sim  m^{-1} $ the scattering region $0<x\lesssim m^{-1}$.

At zero order in $h$, these amplitudes should be related by the scattering condition \cite{Del08}
\be \label{SCq}
C_k(\beta_n)= \{  S_2(2\beta_n)[(q-1)\delta_{k,0}-1]+S_3(2\beta_n)  \}\, C_k^*(\beta_n) .
\ee
The explicit form of the factor in the curly brackets in the right-hand side is:
\bea
\fl (q-2)S_2(\theta)+S_3(\theta)= -e^{{\cal A}(\theta)}
\frac{ \sinh\left[{\lambda}(\frac{\rmi\pi}{3}+\theta) \right] }{ \sinh\left[{\lambda}(\frac{\rmi\pi}{3}-\theta) \right] }
\,\frac{ \sinh[\lambda(\rmi\pi+\theta) ]}{\sinh[\lambda(\rmi\pi-\theta)]}\,\,, \quad {\rm for }\quad k=0, \label{Sq2}\\
\label{Sodd}
\fl S_3(\theta)-S_2(\theta)= -e^{{\cal A}(\theta)}
\frac{ \sinh\left[{\lambda}(\frac{\rmi\pi}{3}+\theta) \right] }{ \sinh\left[{\lambda}(\frac{\rmi\pi}{3}-\theta) \right] }
, \quad {\rm for }\quad k=1,\ldots,q-2.
\eea
Equations (\ref{Ck}), (\ref{SCq}), and (\ref{Sq2}) lead to the semiclassical quantization condition
\be \label{sm0}
\fl\sinh(2\beta_{n})-2\,\beta_{n}=  \left[
2\pi \left(n-\frac{1}{4}\right)+ \rmi {\cal A}(2 \beta_{n})
+2 \alpha_1(2\beta_n)+2 \alpha_2(2\beta_n)\right]\zeta + O(\zeta^2),
\ee
where
\be
\fl \alpha_1 (\theta)=\arctan\left[ \tanh (\lambda  \theta)  \cot(\pi\lambda) \right],\quad
\alpha_2 (\theta)=\arctan\left[ \tanh (\lambda  \theta)  \cot(\pi\lambda/3) \right].
\ee
Equations (\ref{Mn}), (\ref{sm0}) determine the masses of $\pi_0$   for $n\gg1$.

For the multiplet $\pi_k$ , $k=1,...,q-2$, the analogous quantization condition
\be
\fl\sinh(2\beta_{n})-2\,\beta_{n}=  \left[
2\pi \left(n-\frac{1}{4}\right)
 +\rmi {\cal A}(2 \beta_{n})+2 \alpha_2(2\beta_n)
\right]\zeta + O(\zeta^2)    \label{sm1}
\ee
follows from equations (\ref{Ck}), (\ref{SCq}), and (\ref{Sodd}).

For $q=4$,  the masses of $\pi_0$ and  $\pi_k$, $k=1,...,q-2$ are the same:
\be
\fl \sinh(2\beta_{n})-2\,\beta_{n}=  \left[
2\pi \left(n-\frac{1}{4}\right)+
\rmi {\cal A}(2 \beta_{n})
\right]\zeta  + O(\zeta^2) \quad {\rm for} \quad k=0,...,q-2.
\ee
\subsubsection{Low-energy expansion}
Consider  the leading order of the low-energy expansion $h\to +0$, $n\sim 1$. Functions $ \phi_k (x) $ are  smooth
in  this case and vary on the scale much larger than the correlation length $m^{-1}$.
At large positive $x\gg m^{-1}$, they
should satisfy the differential equation
\be
\left[2m -M_n -\frac{1}{m}\, \partial_x^2 +x \, \Delta {\cal E}\right ]\,\phi_k (x) =0. \label{Air}
\ee
The regular at
$x\to\infty$ solution is given by the Airy function
\be
\phi_k (x)={\rm Ai}(t-t_n),
\ee
where
\be
t=  (m\,\Delta {\cal E})^{1/3}\, x  ,\quad t_n= \frac{(M_n-2 m)\, m^{1/3}}{(\Delta {\cal E})^{2/3} }.
\ee
For noninteger $\lambda$ equations  (\ref{Sq2}), (\ref{Sodd}) yield
\bea
&&\lim_{\theta\to 0}\{ S_2(\theta)[(q-1)\delta_{k,0}-1]+S_3(\theta) \}=-1,\\
&&\lim_{\theta\to 0}\{S_3(\theta)-S_2(\theta)\}=-1.
\eea
This implies the fermionic boundary condition for the wave function ${\phi_k(0)=0}$
for $k=0,...,q-2$ leading  to the same low-energy mass spectrum (\ref{lMk}) for all mesons $\pi_k$.
\subsection{Weak coupling expansion for $q=3$}
A separate consideration is needed at $q=3$.
Relations   (\ref{s2}) - (\ref{Aex})  reduce in this case to
\bea \label{S2q3}
&&\lambda=1,\\
&& S_2(\theta)=-\frac{ {\sinh}(\theta+\rmi\pi/3) }{{\rm sinh}(\theta-\rmi\pi/3)}\,e^{ {\cal A}(\theta) }, \quad S_3(\theta)=0,\\
&&{\cal A}(\theta)=\int_0^\infty\frac{\rmd x}{x}\,
\frac{2 \sinh (x/3)}
{\sinh x}\,\sinh\frac{x\theta}{\rmi\pi}\,.
\eea
The scattering condition (\ref{SCq}) takes now the form
\be \label{SC}
C_k(\beta_n)=(-1)^k \,S_2(2\beta_n)\, C_k^*(\beta_n).
\ee
In the low-energy case, the rapidities of quarks are small, $\theta_{12}\ll 1$, and
\be
S_2(\theta_1-\theta_2)\approx S_2(0)=1.
\ee
Thus,  the boundary conditions  for equation (\ref{Air}) should be
bosonic for $ \phi_0(x) $
\be
 \phi_0'(0) =0 ,
\ee
and fermionic for $ \phi_1(x) $
\be
 \phi_1(0)=0 .
\ee
Therefore, whereas the mass spectrum of  $\pi_1$  in the low-energy region is still given by
equation (\ref{lMk}),
the masses of $\pi_0$ are described now by
\be \label{lM+}
\frac{M_n}{m}-2= \zeta^{2/3}  z_n'+O(\zeta^{4/3}),
\ee
with $(-z_n')$ being the zeros of the first derivative of the Airy function: ${\rm Ai}'(-z_n')=0$,
$n=1,2,\ldots$.

It is tempting to associate this peculiar behaviour of the low-energy $\pi_0$ mass spectrum 
in the $q=3$ case with the presence of the $B$-meson   in the zero-field PFT \cite{CZ92}. 
At $ q=3$ and $h \to+0$, its mass $M_B$ lies exactly at the edge point $2 m$ of the $\pi$-meson spectra.

At large $n\gg 1$, the zeros of the Airy function and its derivative behave as \cite{AbrSt}
\be
z_n\approx\left[\frac{3\pi (4n-1)}{8}\right]^{2/3}, \quad
z_n'\approx\left[\frac{3\pi (4n-3)}{8}\right]^{2/3}.
\ee

Combining these asymptotics with (\ref{lMk}), (\ref{lM+}) and (\ref{Mn}), we get
\bea \label{cr}
\frac{(2\beta_{n})^3}{3!}\approx   2\pi\left(n-\frac{3}{4}\right)\zeta,\quad {\rm for} \; k=0, \\
\frac{(2\beta_{n})^3}{3!}\approx  2\pi\left(n-\frac{1}{4}\right) \zeta ,
\quad {\rm for} \; k=1 \label{cr1}
\eea
at  $n\gg 1$.

In the semiclassical region, quantization conditions (\ref{sm0}), (\ref{sm1}) reduce at $q=3$ to the form
\be
\fl \sinh(2\beta_{n})-2\,\beta_{n}=  \Bigg[
2\pi \left(n-\frac{3}{4}\right)+2 \arctan\left( \frac{\tanh 2 \beta_{n}  }{\sqrt{3}} \right)
+\rmi {\cal A}(2 \beta_{n})
\Bigg]\zeta + O(\zeta^{4/3})
\ee
for   $k=0$, and
\be
\fl \sinh(2\beta_{n})-2\,\beta_{n}=\Bigg[
2\pi \left(n-\frac{1}{4}\right)+2 \arctan\left( \frac{\tanh 2 \beta_{n}  }{\sqrt{3}}  \right)
+\rmi {\cal A}(2 \beta_{n})
\Bigg]\zeta + O(\zeta^{4/3}) \nonumber
\ee
for $k=1$.
At $\beta_{n}\to 0$ these relations agree with (\ref{cr}), (\ref{cr1}).
\section{Meson masses at $h\to-0$ \label{hneg}}
At negative magnetic field orientated along the $q$-th direction, the kinks $K_{\alpha q}$ and $K_{q\beta}$
interpolating between the true and the false vacua become confined,  while the
kinks $K_{\alpha\beta}$ connecting two true vacua remain stable. Coupling  two attracting kinks into bound states,
one could construct the meson states both in the topological charged and topological neutral sectors.

In the topological charged sector, the meson state $\pi_{\alpha\beta}(0)$ in the 'transport' region
$x_2-x_1> a/m$, with some constant $a\gg1$, can be written as
\be
\int_{-\infty}^\infty \rmd x_1 \int_{x_1+a/m}^\infty \rmd x_2\,
|K_{\alpha q}(x_1)K_{q\beta}(x_2)\rangle \,\psi_{\alpha\beta} (x_2-x_1),
\ee
where the meson wave function $\psi_{\alpha\beta} (x)$ should satisfy equation (\ref{qc})
with the string tension
\be
\Delta {\mathcal E}=|h|( \langle 0_q| \sigma_q |0_q\rangle- \langle 0_\alpha | \sigma_q|0_\alpha\rangle)=
\frac{q}{q-1}|h| v, \quad \alpha\ne q.
\ee
However, the two kinks can become deconfined after the scattering process
\be \label{ch}
 K_{\alpha q}(\theta_1)K_{q\beta}(\theta_2) \to K_{\alpha \gamma} (\theta_2) K_{\gamma \beta}(\theta_1),
 \quad\quad\gamma\ne q,
\ee
characterized by the amplitude $S_0(\theta_{12})$ in (\ref{KSc}). As the result, the topologically
charged mesons $\pi_{\alpha\beta}(P)$ are unstable already in the leading order in $h$ for $q>3$.

On the other hand, the decay channel (\ref{ch}) is evidently closed for $q=3$. For the state  $K_{1 3}(\theta_1)K_{32}(\theta_2)$,
the remaining scattering process
\be
 K_{13}(\theta_1)K_{32}(\theta_2) =S_1(\theta_{12}) K_{13} (\theta_2) K_{32}(\theta_1),
\ee
for the state  $K_{1 3}(\theta_1)K_{32}(\theta_2)$ is characterized by  amplitude (\ref{S1}),
which reduces at $q=3$ to the form
\be
S_1(\theta)=-e^{{\mathcal A}(\theta)}.
\ee
Reproducing with minimal changes calculations described in Section \ref{Sect3}, we obtain the masses
of the topologically charged mesons $\pi_{12}$ for $q=3$.
 For large $n\gg1$, they are described  by the semiclassical quantization condition
\be \label{sm3}
\sinh(2\beta_{n})-2\,\beta_{n}=  \left[
2\pi \left(n-\frac{1}{4}\right)+ \rmi {\cal A}(2 \beta_{n})\right]\zeta + O(\zeta^2),
\ee
and equation(\ref{Mn}).   For small $n$, $n\ll \zeta^{-1}$ the  masses of  $\pi_{12}$ are described by
equation (\ref{lMk}).
Note, that $\zeta=\Delta{\mathcal E}/m^2\sim |h|$ at negative $h$.

In the topological neutral sectors, mesons $\pi_{\alpha\alpha}$ at $q\ne 4$ can easily decay due to decoupling of kinks in the
process
\be \label{ch2}
 K_{\alpha q}(\theta_1)K_{q\alpha}(\theta_2) \to K_{\alpha \gamma} (\theta_2) K_{\gamma \alpha}(\theta_1),
 \quad\quad\gamma\ne q.
\ee
In zero order in $h$, the scattering amplitude $S_2 (\theta_{12}) $  of this channel  is given by equation (\ref{s2}).

The kink decoupling  is hindered at $q=4$, since $S_2(\theta_{12})$ vanishes in this case.
As the result, commutation relation (\ref{s3}) for the mutual scattering of quarks $K_{\alpha q}(\theta_1)$, $K_{q\alpha}(\theta_2)$
reduces at $q=4$  to the form
\be \label{ch3}
 K_{\alpha 4}(\theta_1)K_{4\alpha}(\theta_2) = S_3(\theta_{12}) K_{\alpha 4} (\theta_2) K_{4 \alpha}(\theta_1),
\ee
where $\alpha=1,2,3$, and
\be
S_3(\theta)=-e^{ {\mathcal A}(\theta) }.
\ee
Accordingly, the masses of $\pi_{\alpha\alpha}$ are described at $q=4$ by relations (\ref{lMk}) and
(\ref{sm3}) in the leading order in $h$. Note, that the function ${\mathcal A}(\theta)$ depends also on parameter $q$ through its
(not indicated explicitly) dependence on  $\lambda$, see equation (\ref{Aex}). It is natural to expect, that   channel (\ref{ch2})
opens in higher orders in $h$ making unstable  all the mesons in the topologically neutral sector
 at $h\to-0$ and  $q=4$.
\section{Bethe-Salpeter equation \label{sect:BS}}
The heuristic approach applied in this paper is based on assumptions adopted in Subsection \ref{Genq}.
We  postulate the simple form (\ref{Vi}) of the interaction between quarks at large distances, 
take the quark dispersion law in the form (\ref{eii}), and apply
the boundary condition (\ref{SCq}) originating from the scattering matrix at $h=0$.
Though this  procedure seems to be sufficient for  determining the leading order of the meson masses at $|h|\to 0$,
it is desirable to have a more systematic theory suitable for verification of the adopted in Subsection \ref{Genq}  
assumptions, and  for calculation of  higher order corrections to $M_n$. 
In the IFT, an efficient technique 
based on the Bethe-Salpeter equation has been developed by Fonseca and Zamolodchikov \cite{FZ06}. 
In this Section we describe, how a similar Bethe-Salpeter equation for the PFT can be derived. 
We hope, that it will be used in the future for more consequent  calculation of the 
weak-coupling expansion of the meson masses in the PFT. 

The meson energy spectra $E(P)$ are determined by the eigenvalue problem (\ref{en}), (\ref{mom}),
which we rewrite as
\begin{eqnarray}  \label{en1}
\hat{\mathcal H}_0 \,|\pi(P)\rangle -h \int_{-\infty}^\infty \rmd x \,\sigma_q(x)|\pi(P)\rangle  =
\left[E(P) +E_{vac} \right]\,|\pi(P)\rangle,\\
\hat{P} \,|\pi(P)\rangle ={P} \,|\pi(P)\rangle. \nonumber
\end{eqnarray}
Here the Hamiltonian $\hat{\mathcal H}_0$ corresponds to the integrable zero-field PFT in the ordered 
phase. Integrability of PFT at $h=0$ implies, in particular, that the Hamiltonian 
${\mathcal H}_0$ (together with the momentum operator $\hat{P}$) can be diagonalized by the multi-kink
states
$   |K_{\underline{\alpha}} (\underline{p})\rangle $:
\begin{eqnarray} \label{bas}
&&\hat{\mathcal H}_0 |K_{\underline{\alpha}} (\underline{p})\rangle=\left[\omega(p_1) +\ldots+\omega(p_n)\right]\,
|K_{\underline{\alpha}} (\underline{p})\rangle ,\\
&&\hat{P} \, |K_{\underline{\alpha}} (\underline{p})\rangle= (p_1 +\ldots+p_n)
 \, |K_{\underline{\alpha}} (\underline{p})\rangle , \label{mm}
\end{eqnarray}
where 
\begin{eqnarray}  \label{bas1}
|K_{\underline{\alpha}} (\underline{p})\rangle= |K_{\alpha_0,\alpha_1}(p_1)K_{\alpha_1,\alpha_2}(p_2) 
\ldots K_{\alpha_{n-1},\alpha_n}(p_n)\rangle \\
= \frac{|K_{\alpha_0,\alpha_1}(\theta_1)K_{\alpha_1,\alpha_2}(\theta_2) \nonumber
\ldots K_{\alpha_{n-1},\alpha_n}(\theta_n)\rangle }{\left[\omega(\theta_1)\ldots\omega(\theta_n)\right]^{1/2}}  ,     
\end{eqnarray}
$\alpha_{j+1}\ne \alpha_j$, the kink momenta are ordered as $\infty>p_1>p_2>\ldots>p_n>-\infty$, and 
$\theta_j={\textrm {arcsinh}}(p_j/m)$ are the corresponding rapidity variables. 
The kink states in the momentum and rapidity bases are normalized as 
\begin{eqnarray*}  \langle K_{\beta,\alpha}(p)|K_{\alpha,\beta}(p')\rangle=2\pi\delta(p-p'), \\
\langle K_{\beta,\alpha}(\theta)|K_{\alpha,\beta}(\theta')\rangle=2\pi\delta(\theta-\theta').
\end{eqnarray*}

Note, that the particle sector of the PFT at $3< q\le 4$, $h=0$, contains also the 
topologically-neutral kink-antikink bound states   $B(p)$, see \cite{CZ92}. 
Of course, the mesons $B(p)$ can appear together with kinks in the asymptotical
in- and out-states at $h=0$. We do not display the mesons $B(p_j)$ explicitly in  formulas  (\ref{bas})-(\ref{basi})
  just to avoid too cumbersome notations. 

Let us now turn to equations (\ref{en1}), concentrating on the case of a positive magnetic field, $h>0$. 
Then, the meson vector $ |\pi(P)\rangle $ being a 
topologically-neutral state in the sector $q$ should admit the expansion
\begin{eqnarray} \label{basi}
|\pi(P)\rangle =\sum_{n=2}^\infty\,\,\sum_{\alpha_1,\ldots,\alpha_{n-1}\ne q}
\int_{\infty>p_1>\ldots>p_n>-\infty}\frac{\rmd p_1\ldots \rmd p_n}{(2\pi)^{n}}\\
\fl \cdot | K_{q,\alpha_1}(p_1)K_{\alpha_1,\alpha_2}(p_2) \ldots K_{\alpha_{n-1},q}(p_n)    \rangle
 \langle 
K_{q,\alpha_{n-1}}(p_n)\ldots K_{\alpha_2,\alpha_1}(p_2) K_{\alpha_{1},q}(p_1)|\pi(P) \rangle \,.\nonumber 
\end{eqnarray}
In complete analogy with the IFT, the two-quark approximation is based on the 
assumption, that at $h\to+0$, the first term with $n=2$ dominates in the infinite sum over $n$ in (\ref{basi}).
Accordingly, in the two-quark approximation one replaces the exact eigenvalue problem (\ref{en1})
by its projection onto the two-quark subspace $ {\mathbf H}_2^{(q)} $ spanned by the basis 
$|K_{q,\alpha}(p_1)K_{\alpha,q}(p_2)\rangle$,
with $\alpha\ne q$ and $p_1>p_2$,
\begin{eqnarray} \label{en2}
{\mathcal H}_0 \,|\tilde{\pi}(P)\rangle -h \int_{-\infty}^\infty \rmd x \,
{\mathcal P}_2^{(q)} \sigma_q(x)|\tilde{\pi}(P)\rangle  =
\left[\tilde{E}(P) +\tilde{E}_{vac} \right]\,|\tilde{\pi}(P)\rangle,  \\
\hat{P} \,|\tilde{\pi}(P)\rangle ={P} \,|\tilde{\pi}(P)\rangle, \nonumber
\end{eqnarray}
where $|\tilde{\pi}(P)\rangle\in {\mathbf H}_2^{(q)}$, and ${\mathcal P}_2^{(q)}$ is the orthogonal projector on
$ {\mathbf H}_2^{(q)} $. 
Tildes distinguish solutions of (\ref{en2})
from those of the exact eigenvalue problem (\ref{en1}).
The meson state in this approximation is characterized by 
the two-quark wave function $ \Psi_\alpha (p_1,p_2) $,
\begin{equation}
 \Psi_\alpha (p_1,p_2)=\langle K_{q,\alpha} (p_2)K_{\alpha,q}(p_1)| \tilde{\pi}(P)\rangle. 
\end{equation}
This relation defines $ \Psi_\alpha (p_1,p_2) $ in the domain $\infty>p_1>p_2>-\infty$. 
Continuation into the whole plane $-\infty<p_1,p_2<\infty$ provided by the 
Faddeev-Zamolodchikov commutation relations (\ref{FZ}), (\ref{KS}) yields
\begin{eqnarray} \label{psisym}
\Psi_\alpha (p_2,p_1) =\sum_{\beta\ne q}S_{qq}^{\beta\alpha}(\theta_1-\theta_2)\Psi_\beta(p_1,p_2)\\
=       S_3(\theta_1-\theta_2)\,\Psi_\alpha (p_1,p_2)+
S_2 (\theta_1-\theta_2) \sum_{\beta\ne\alpha,q} \Psi_\beta(p_1,p_2) . \nonumber
\end{eqnarray}
Then, equation (\ref{en2}) takes the form
\begin{eqnarray}  \label{inte}
\fl \left[\omega(p_1)+\omega(p_2)- \tilde{E}(P)  \right] \Psi_\alpha (p_1,p_2) =\tilde{E}_{vac}\Psi_\alpha (p_1,p_2)+
\frac{h}{2}\int_{-\infty}^\infty \rmd x \int_{-\infty}^\infty \frac{\rmd p_1' \,\rmd p_2'}{(2\pi)^2}\\
\fl \cdot \exp[\rmi x(p_1'+p_2'-p_1-p_2)]\,
 \sum_{\beta=1}^{q-1}  \langle K_{q,\alpha}( p_2)K_{\alpha,q}( p_1)| 
\sigma_q(0)|K_{q,\beta}(p_1')K_{\beta,q}(p_2')\rangle \, \Psi_\beta(p_1',p_2').
\nonumber
\end{eqnarray}
The matrix element in the 
integral kernel of this equation is the generalized formfactor of the 
magnetization operator in the momentum representation. It is well known, that 
such formfactors have the so-called {\it kinematic singularities}
at coinciding in- and out- momenta of particles. 
Let us extract a part of these singularities which are  contained in the 
disconnected 'direct propagation' terms \footnote{A detailed analysis of
 disconnected terms in the integrals containing formfactors in integrable quantum field theories 
in a finite volume has been done by Pozsgay and Tak{\'a}cs \cite{PozTak08}.},  
\begin{eqnarray}
\langle K_{q,\alpha}( p_2)K_{\alpha,q}( p_1)| 
\sigma_q(0)|K_{q,\beta}(p_1')K_{\beta,q}(p_2')\rangle =
{\mathcal G}_{\alpha\beta}(p_2,p_1|p_1',p_2') \\
+ 4\pi^2  \langle \sigma_q\rangle_q
\left[
\delta_{\alpha_\beta} \delta(p_1-p_1')\delta(p_2-p_2')
+S_{qq}^{\beta\alpha}(\theta_1'-\theta_2') \delta(p_1-p_2')\delta(p_2-p_1')
\right].  \nonumber
\end{eqnarray}
Substitution of the second line of the  above formula  into the right-hand side of (\ref{inte}) yields
\bea \nonumber
\fl  \tilde{E}_{vac} \Psi_\alpha (p_1,p_2)+ 4\pi^2h \,\langle \sigma_q\rangle_q  
\int_{-\infty}^\infty \rmd x\int_{-\infty}^\infty 
\frac{\rmd p_1' \,\rmd p_2'}{(2\pi)^2}\,
\delta(p_1-p_1')\delta(p_2-p_2')   \Psi_\alpha(p_1',p_2')\\
\fl\cdot \exp[\rmi x(p_1+p_2-p_1'-p_2')]
= \left(\tilde{E}_{vac} + h  \langle \sigma_q\rangle_q \,\,\int_{-\infty}^\infty \rmd x  \right) \Psi_\alpha(p_1,p_2) .
\label{vac}
\eea
In deriving of the left-hand side of  (\ref{vac}),  we have taken into account the symmetry
(\ref{psisym}). 
It is natural to expect, that the two 
divergent in the thermodynamic limit terms in the brackets in the right-hand side of 
(\ref{vac}) cancel each other. Really, in a finite system of the length $L$, 
the shift of the vacuum energy in presence of a positive magnetic field $h$
is (in the two-quark approximation) $ \tilde{E}_{vac} = -h L \langle \sigma_q\rangle_q  $, and the integral  
$\int \rmd x $ should produce the length of the system $L$ .

After cancelation of the  infinite terms described above, the right-hand side of (\ref{inte})
becomes well defined in the thermodynamic limit, and we can safely
perform in it integration in $x$.  Then,    equation (\ref{inte}) takes the 	final form
in the  variables $p=(p_1-p_2)/2$, $p'=(p_1'-p_2')/2$: 
\begin{eqnarray}  \label{BSPotts}
\left[\omega(P/2+p)+\omega(P/2-p)- \tilde{E}(P)  \right] \Phi_\alpha(p;P) \\=
\frac{h}{2} \int_{-\infty}^\infty \frac{\rmd p'}{2\pi}
 \sum_{\beta=1}^{q-1}  G_{\alpha\beta}(p|p';P)\, \Phi_{\beta}(p';P).
\nonumber
\end{eqnarray}
Here the meson wave function $\Phi_{\alpha}(p;P)$ and the  kernel 
 $G_{\alpha\beta}(p|p';P)$ are defined as
\begin{eqnarray}\label{Phii}
\Psi_\alpha(p_1,p_2)=2\pi \delta(p_1+p_2-P)\Phi_\alpha(p_1-P/2),\\
 G_{\alpha\beta}(p|p';P) = {\mathcal G}_{\alpha\beta}(P/2-p,P/2+p|P/2+p',P/2-p').
\end{eqnarray}
The transformation law for the function  $\Phi_\alpha(p;P)$ under the reflection
$p\to-p$ can be read from (\ref{psisym}) and (\ref{Phii}). 

Equation (\ref{BSPotts}) gives generalization of the IFT Bethe-Salpeter equation 
to the $q$-state PFT with $2<q\le4$. Though equation (\ref{BSPotts}) and its derivation look 
very similar to those in the IFT (see Section 3 and equation (3.11) in reference  \cite{FZ06}),
an important difference should be pointed out. Derivation of the Bethe-Salpeter equation in the IFT
is based on the free-fermionic basis, since the IFT describes noninteracting particles at $h=0$. 
In contrast, the quarks in the PFT at $q>2$ and $h=0$ strongly interact at small distances $\sim m^{-1}$. 
The basis states 
(\ref{bas1}), which diagonalize the PFT zero-field Hamiltonian $\hat{\mathcal H}_0$, 
can be treated as its $n$-quark 
'stationary scattering states'. Corresponding wave functions in the coordinate representation 
look like a superposition of plane waves only if the distances between quarks are large compared with the 
interaction radius. In particular, the wave function for the basis state 
 $| K_{q,\alpha} (p_1)K_{\alpha,q}(p_2)\rangle$ reads as 
\be
\fl \psi_\beta(x_1,x_2)=\delta_{\alpha \beta} \exp[\rmi (p_1 \,x_1+p_2 \,x_2)]+ S_{qq}^{\alpha\beta}(\theta_1-\theta_2)
\exp[\rmi (p_2 \,x_1+p_1\, x_2)], 
\ee
if $x_2-x_1\gg m^{-1}$. Here $p_1>p_2$ is implied for the basis state, and 
$x_1$ and $x_2$ denote the spacial coordinates of the left and right quarks, respectively, 
$-\infty<x_1<x_2<\infty$. However,  in the interaction region  $0<x_2-x_1\lesssim m^{-1}$, 
the motion corresponding to the state $| K_{q,\alpha} (p_1)K_{\alpha,q}(p_2)\rangle$ is much more complicated, 
and it "cannot be treated in terms of the wave function of a finite number of variables 
(because the virtual pair creation is possible)"  \cite{ZZ79}. Therefore, one should not understand
the term 'the two-quark approximation'   in a literal sense in the 
$q$-state PFT with $q\ne2$.  

The kernel  $G_{\alpha\beta}(p|p';P)$ of the Bethe-Salpeter equation (\ref{BSPotts}) is simply related with the 
matrix element of  the magnetization operator between the two-kink states in the rapidity basis,
\begin{eqnarray} \label{ME}
 \langle K_{q,\alpha}( \theta_2)K_{\alpha,q}( \theta_1)| 
\sigma_q(0)|K_{q,\beta}(\theta_1')K_{\beta,q}(\theta_2')\rangle. \\
=\left[ \omega(p_1) \omega(p_2)\omega(p_1')\omega(p_2')  \right]^{1/2}\langle K_{q,\alpha}( p_2)K_{\alpha,q}( p_1)| 
\sigma_q(0)|K_{q,\beta}(p_1')K_{\beta,q}(p_2')\rangle.
\nonumber
\end{eqnarray}
The latter matrix element can be expressed with the help of the crossing relations in terms of the 
four-kink elementary formfactor of the magnetization operator
\be \fl \label{FF4}
F_{\alpha_0 \alpha_1 \alpha_2\alpha_3 \alpha_4}^{\sigma_q}(\theta_1,\theta_2,\theta_3,\theta_4) =
\langle 0_{\alpha_0}|\sigma_q(0)| K_{ \alpha_0,\alpha_1 }(\theta_1)K_{ \alpha_1,\alpha_2 }(\theta_2)
K_{ \alpha_2,\alpha_3 }(\theta_3)K_{ \alpha_3,\alpha_4}(\theta_4)\rangle,
\ee
with $\alpha_4=\alpha_0$.
Formfactors are the central objects in the formfactor bootstrap approach in the two-dimensional
integrable quantum field theories \cite{Berg79,KW78,Sm92}. The $n$-particle formfactors
 are subject to a set of equations (axioms), 
which often allow to calculate them exactly (for a review see \cite{Sm92}).  
Unfortunately, the explicit expressions for all $n$-kink formfactors in the $q$-state PFT is  known only 
in the Ising case $q=2$. Delfino and Cardy \cite{DC98} obtained  the 
two-kink formfactors $\langle 0_\alpha|\sigma_\gamma(0)| K_{ \alpha,\beta }(\theta_1)K_{ \beta,\alpha }(\theta_2)
\rangle$ of the magnetization operators for $q=3,4$.   

Calculation of the four-kink formfactors (\ref{FF4}) would be crucial for determining the  kernel 
$G_{\alpha\beta}(p|p';P)$ of equation (\ref{BSPotts}). 
This kernel should be singular at $p=\pm p'$ due to the kinematic singularities
of the function ${\mathcal G}_{\alpha\beta}(p_2,p_1|p_1',p_2')$. We expect, that the leading singularity
of $G_{\alpha\beta}(p|p';P)$  at $p\to p'$ has the form
\be
\delta_{\alpha\beta}\left( \langle\sigma_q\rangle_q  -
\langle\sigma_q\rangle_\alpha\right)
\left[\frac{1}{(p-p'+\rmi 0)^2} + \frac{1}{(p-p'-\rmi 0)^2}\right].
\ee
This term should produce after the Fourier transform the long range 
attractive potential $\Delta{\mathcal E}|x| $ between the quarks. On the
other hand, the regular at $p=\pm p'$ part of  $G_{\alpha\beta}(p|p';P)$
should describe the change in their short-range interaction induced by the magnetic field.

As we know from the perturbative solution of the Bethe-Salpeter equation in the IFT \cite{FZ06,Rut09}, 
the regular (short-range) part of the integral kernel contributes into the meson masses $M_n(h)$ only
in the {\it second} order in $h$. Physically, the additional factor $h$ reflects that 
the two quarks bound in the meson spend at $h\to 0$ almost all the time at large distances. 
They only rarely come up in the scattering region $x_2-x_1\sim m^{-1}$ where the  $h$-order
correction to the 
short-range interaction due to the regular part of $G_{\alpha\beta}(p|p';P)$
should be taken into account.  
\section{Conclusion \label{Conc}}
We extended the heuristic perturbative approach, which was  originally developed \cite{FZ06, 
FonZam2003, Rut05}
 for calculation of the meson masses in the weak confinement regime in the Ising field theory,
to the $q$-state Potts field theory with  $2<q\le 4$ in presence of a weak magnetic field $h$.
Though the latter model is integrable at $h=0$,  the kinks ('quarks') remain to be interacting particles
 at $h=0$. We  have calculated the masses $M_n(h)$ of the mesons  in the PFT at $T<T_c$ in the leading order in the
 weak magnetic field $|h|\to 0$ both in the low energy, and semiclassical cases. 
The mesons with nonzero topological charge were predicted for the $3$-state PFT 
in the ordered phase at $h<0$.
The Bethe-Salpeter equation is derived for the $q$-state PFT with $2<q\le 4$, which 
generalizes the analogous equation known in the IFT. This equation could provide a more firm basis
for the theory, if the explicit expressions for the four-kink formfactors of the magnetization
operator would be found.

After the first version of this articles had appeared as a preprint, Lepori, T\'oth and Delfino 
\cite{LTD09} presented the results of their numerical investigations of the 
particle  spectra in the $3$-state PFT in a wide 
range of the magnetic fields $h$ and temperatures $\tau$ by means of the Truncated Conformal 
Space Approach (TCSA) \cite{YuZam90}. They confirmed the qualitative picture of confinement developed 
in reference  \cite{Del08}, and were able to partly confirm our analytical predictions (\ref{lMk}) 
and (\ref{lM+}) for the low-energy part of the 
 meson spectra at $h\to+0$. Reported in reference \cite{LTD09} magnetic field 
dependence of the masses of five lightest even ($i=0$) and odd ($i=1$)  mesons  
at moderately large magnetic fields were described by formula 
\be
M_n^{(i)}=2 m + c_n^{(i)} h^{\alpha},
\ee
with
\be
\alpha\approx 0.7, \quad \frac{c_1^{(1)}}{c_1^{(0)}} \approx 2.
\ee
These values are in reasonable agreement with the numbers 
\be
\alpha=\frac{2}{3}, \quad   \frac{c_1^{(1)}}{c_1^{(0)}}  =\frac{z_1}{z_1'}\approx 2.3,
\ee
 following from (\ref{lMk}) 
and (\ref{lM+}).
However, for a complete numerical verification 
one needs to increase the accuracy of the TCSA calculations at small $|h|$. 

The obtained results could be developed  in several directions. First, it is straightforward to
extend them to the wide class of  models exhibiting confinement, which are integrable but not
free at zero 'magnetic field', see reference \cite{MusTak09}. Regarding the Potts field theory,
one can try to find from the
Bethe-Salpeter equation (\ref{BSPotts}) corrections to the meson masses at small $|h|$,  and to study the decay
mechanisms for unstable mesons in the higher orders in $|h|$. 

It is remarkable \cite{FZ06}, that in the  IFT, the  Bethe-Salpeter equation  reproduces with reasonable accuracy
the mesons masses  not only
in the limit $h\to 0$, but also at finite, and even at large values of the magnetic field $h$.
If this situations holds also for the PFT,  equation (\ref{BSPotts})
could be useful for nonperturbative calculations of the meson spectra in the PFT  at finite 
magnetic fields.
To achieve progress in all these directions, it is essential to find explicit expressions
for the $n$-kink formfactors of the magnetization operator in the PFT.

One more interesting open problem
is to determine in the PFT  the masses of 'baryons' consisting of three quarks \cite{Del08}. 

We close with the following remark. For the sake of simplicity, we calculated in Sections \ref{Sect3} and \ref{hneg} the energy
of a meson which has zero momentum, i.e. analyzed the problem in the meson rest frame.
It is straightforward to modify calculations to the case of a generic frame, and to check
(in the leading order in $h$), that the meson dispersion law $E(P)$ really has the relativistic form (\ref{relM}),
as one should expect.
\noindent
\ack
I thank G. Delfino for interesting discussion.
I am thankful for hospitality to the Abdus Salam International Centre for Theoretical
Physics, Trieste, where the first version of this article was finished.\newline
This work is supported by  the Belarusian Republican Foundation for Fundamental Research.
 
\section*{Referencses}


\begin{thebibliography}{30}

\bibitem{McCoy78}
B.~M. McCoy and T.~T. Wu.
\newblock Two-dimensional {I}sing field theory in a magnetic field: Breakup of
  the cut in the two-point function.
\newblock {\em Phys Rev. D}, 18(4):1259--1267, 1978.

\bibitem{FZ06}
P.~Fonseca and A.~B. Zamolodchikov.
\newblock Ising spectroscopy ${{\rm I}} $: Mesons at ${T}<{T}_c$, 2006.
\newblock arXiv:hep-th/0612304.

\bibitem{Tsv04}
M.~J. Bhaseen and A.~M. Tsvelik.
\newblock Aspects of confinement in low dimensions, 2004.
\newblock arXiv:cond-mat/0409602.

\bibitem{DelMus98}
G.~Delfino and G.~Mussardo.
\newblock Non-integrable aspects of the multi-frequency sine-{G}ordon model.
\newblock {\em Nucl. Phys. B}, 516(3):675--703, 1998.
\newblock (arXiv:hep-th/9709028).

\bibitem{Del08}
G.~Delfino and P.~Grinza.
\newblock Confinement in the $q$-state {P}otts field theory.
\newblock {\em Nucl. Phys. B}, 791(3):265--283, 2008.
\newblock (arXiv:hep-th:0706.1020).

\bibitem{Mus07}
G.~Mussardo.
\newblock Kink confinement and supersymmetry.
\newblock {\em JHEP}, 08:03, 2007.
\newblock (arXiv:hep-th/0706.2546).

\bibitem{Mus08}
L.~Lepori, G.~Mussardo, and G.~Zs. T\'oth.
\newblock The particle spectrum of the tricritical {I}sing model with spin
  reversal symmetric perturbations.
\newblock {\em J. Stat. Mech.}, P09004, 2008.
\newblock (arXiv:hep-th/0806.4715).

\bibitem{MusTak09}
G.~Mussardo and G.~Tak\'acs.
\newblock Effective potentials and kink spectra in non-integrable perturbed
  conformal field theories.
\newblock {\em J. Phys. A}, 42(30):304022, 2009.
\newblock (arXiv:hep-th/0901.3537).

\bibitem{FonZam2003}
P.~Fonseca and A.~B. Zamolodchikov.
\newblock Ising field theory in a magnetic field: Analytic properties of the
  free energy.
\newblock {\em J. Stat. Phys.}, 110(3-6):527--590, 2003.
\newblock (arXiv:hep-th/0112167).

\bibitem{Rut05}
S.~B. Rutkevich.
\newblock Large-$n$ excitations in the ferromagnetic {I}sing field theory in a
  weak magnetic field: Mass spectrum and decay widths.
\newblock {\em Phys. Rev. Lett.}, 95(25):250601, 2005.
\newblock (arXiv:hep-th/0509149).

\bibitem{Rut09}
S.~B. {Rutkevich}.
\newblock Formfactor perturbation expansions and confinement in the {I}sing
  field theory.
\newblock {\em J. Phys. A}, 42(30):304025, 2009.
\newblock (arXiv:cond-mat/0901.1571).

\bibitem{CZ92}
L.~Chim and A.~B. Zamolodchikov.
\newblock Integrable field theory of the $q$-state {P}otts model with $0 < q <
  4$.
\newblock {\em Int. J. Mod. Phys. A}, 7(21):5317--5336, 1992.

\bibitem{KasFor69}
P.~W. Kasteleyn and E.~M. Fortuin.
\newblock {\em J. Phys. Soc. Japan. Suppl.}, 26:11, 1969.

\bibitem{KasFor72}
P.~W. Kasteleyn and E.~M. Fortuin.
\newblock On the random-cluster model : I. {I}ntroduction and relation to other
  models.
\newblock {\em Physica}, 57(4):536--564, 1972.

\bibitem{Bax}
R.~J. Baxter.
\newblock {\em Exactly solved models in statistical mechanics}.
\newblock Academic Press, London, 1982.

\bibitem{Wu82}
F.~Y. Wu.
\newblock The {P}otts model.
\newblock {\em Rev. Mod. Phys.}, 54(1):235 -- 268, 1982.

\bibitem{Berg79}
B.~Berg, M.~Karowski, and P.~Weisz.
\newblock Construction of {G}reen's functions from an exact {$S$} matrix.
\newblock {\em Phys. Rev. D}, 19(8):2477--2479, 1979.

\bibitem{AbrSt}
M.~Abramowitz and I.~A. Stegun.
\newblock {\em Handbook of Mathematical Functions}.
\newblock Dover, New York, 1965.

\bibitem{PozTak08}
B.~Pozsgay and G.~Tak\'acs.
\newblock Form factors in finite volume {II}: {D}isconnected terms and finite
  temperature correlators.
\newblock {\em Nucl. Phys. B}, 778(3):209--251, 2008.
\newblock (arXiv:hep-th/0706.360).

\bibitem{ZZ79}
A.~B. Zamolodchikov and Al.~B. Zamolodchikov.
\newblock Factorized {$S$}-matrices in two dimensions as the exact solutions of
  certain relativistic quantum field theory models.
\newblock {\em Ann. Phys.}, 120:253--291, 1979.

\bibitem{KW78}
M.~Karowski and P.~Weisz.
\newblock Exact form factors in (1 + 1)-dimensional field theoretic models with
  soliton behaviour.
\newblock {\em Nucl. Phys. B}, 139(4):455--476, 1978.

\bibitem{Sm92}
F.~A. Smirnov.
\newblock Form-factors in completely integrable models of quantum field theory.
\newblock {\em Adv. Ser. Math. Phys.}, 14:1--208, 1992.

\bibitem{DC98}
G.~Delfino and J.~L. Cardy.
\newblock Universal amplitude ratios in the two-dimensional $q$-state {P}otts
  model and percolation from quantum field theory.
\newblock {\em Nucl. Phys. B}, 519:551, 1998.
\newblock (arXiv:hep-th/9712111).

\bibitem{LTD09}
L.~Lepori, G.~Zs. T\'oth, and G.~Delfino.
\newblock Particle spectrum of the $3$-state {P}otts field theory: a numerical
  study.
\newblock {\em J. Stat. Mech.}, P11007, 2009.
\newblock arXiv:hep-th/0909.2192.

\bibitem{YuZam90}
V.~P. Yurov and {Al}.~B. Zamolodchikov.
\newblock Truncated conformal space approach to scaling {L}ee-{Y}ang model.
\newblock {\em Int. J. Mod. Phys. A}, 5(16):3221 -- 3245, 1990.

\end{thebibliography}
\end{document}